\documentclass[english,showpacs,preprintnumbers,amsmath,amssymb,twocolumn]{revtex4}
\usepackage[T1]{fontenc}
\usepackage[latin1]{inputenc}
\usepackage{graphicx}

\makeatletter

\providecommand{\tabularnewline}{\\}

\makeatletter

\usepackage{dcolumn}

\usepackage{bm}

\usepackage{epsfig}

\def\b{\begin{equation}}
\def\e{\end{equation}}

\makeatother

\usepackage{babel}
\makeatother
\begin{document}

\title{Electric quadrupole moment of the $5d^{2}D_{3/2}$ state in $\mathrm{^{171}Yb^{+}}$
: A relativistic coupled-cluster analysis}

\author{K. V. P. Latha$^{1}$}

\email{latha@iiap.res.in}

\author{C. Sur$^{1,2}$ }

\email{csur@astronomy.ohio-state.edu}

\author{R. K. Chaudhuri$^{1}$, B. P. Das$^{1}$ and D. Mukherjee$^{3}$}

\affiliation{$\,^{1}$Non-Accelerator Particle Physics Group, Indian Institute
of Astrophysics, Bangalore - 560 034, India}

\affiliation{$\,^{2}$Department of Astronomy, The Ohio State University, Columbus,
Ohio, 43210, USA}

\affiliation{$\,^{3}$Department of Physical Chemistry, Indian Association for
the Cultivation of Science, Kolkata - 700 032, India}

\date{Last revision : August 15, 2007 : 12:55 hrs : CS}

\begin{abstract}
The electric quadrupole moment for the $5d^{2}D_{3/2}$ state of $\mathrm{^{171}Yb^{+}}$,
has been calculated using the relativistic coupled-cluster method.
Earlier a similar calculation was performed for the 4d $^2D_{5/2}$ state of
$\mathrm{^{88}Sr^{+}}$
which is the most accurate determination to date {[}PRL, \textbf{96},
193001 (2006)]. The present calculation of the electric quadrupole
moment of $\mathrm{^{171}Yb^{+}}$ yielded a value $2.157 ea_{0}^{2}$
where the experimental value is $2.08(11)ea_{0}^{2}$; $a_{0}$ is
the Bohr radius and $e$ the elementary charge. We discuss in this
paper our results in detail for $^{171}{\rm Yb}^{+}$ and highlight the
dominant correlation effects present. We have presented the effect
of inner core excitations and their contribution to the electric quadrupole
moment, which is a property sensitive to regions away from the nucleus. 
\end{abstract}


\maketitle

\section{Introduction}

The frequency of any periodic event like the mechanical oscillation
of pendulum or stable atomic frequencies can be used to define the
unit of time. The frequencies derived from selected atomic resonant
transitions are particularly preferred due to various advantages they
offer compared to mechanical oscillations. They are extremely stable,
accurately measurable and reproducible. Though the cesium atomic clock
frequency \cite{wynands,NIST : clocks} is accurate to 4 parts in $10^{16}$, a variety of atoms
and ions have been proposed as candidates for the next generation
of frequency standards \cite{barwood,hollberg-review}. Of these candidates,
certain ions with forbidden transitions in the optical regime are of special
importance. Trapped and laser cooled single ions and neutral
atoms trapped in optical lattices are currently the leading candidates
for atomic clocks \cite{oskay1,takamoto}. Frequency
standard experiments with trapped ions require considerable skill and ingenuity.
It is a indeed very challenging to measure the clock frequencies to a high degree of
precision. Effects
like the second-order Zeeman, the electric quadrupole shift, etc.
arise from the interaction of the ion with stray fields. It was shown
recently that the systematic effects caused by these shifts will not
limit the accuracy of the optical clock \cite{hg+_prl}.

It was shown earlier that Yb$^{+}$ is one of the suitable candidates
for defining a frequency standard \cite{stenger,Yb+_Q-exp}. Other
candidates like Sr$^{+}$ \cite{bernard,margolis}, Ca$^{+}$ \cite{ca+},
Ba$^{+}$ \cite{ba+}, Hg$^{+}$ \cite{oskay1} etc have also been
considered for setting up the frequency standard. In particular, Yb$^{+}$
is a very versatile candidate, having the clock transitions in the
visible, IR and microwave regions. The transition which is being considered
for the frequency standard experiment is the forbidden electric quadrupole
($E2$) transition between the ground state ($6s\,^{2}S_{1/2},F=0$)
and the metastable excited state ($5d\,^{2}D_{3/2},F=2$).
Precise measurements of the electric quadrupole
moments of these ions have been performed \cite{hg+_prl,Yb+_Q-exp,stenger,barwood-srQ,dube}.
Comparison of the experimental values of electric quadrupole moments
with those calculated theoretically, would serve as excellent tests
of relativistic atomic theories. An earlier calculation of this quantity
for $\mathrm{Sr^{+}}$, using relativistic coupled-cluster theory
yielded a value $\left(2.94\pm0.07\right)ea_{0}^{2}$ \cite{csur-PRL} for the
4d $^2D_{5/2}$ state, which was in very good agreement with it's measured value $\left(2.6\pm0.3\right)$
\cite{barwood-srQ}, where $e$ is the electronic charge and $a_{0}$
is the Bohr radius. This was the most accurate calculation of the
quantity for the 4d $^2D_{5/2}$ state of $\mathrm{Sr^{+}}$ to date and the excellent agreement
of the measured and the calculated values indicates the potential
of the method used. In this paper, the electric quadrupole moment
of $\mathrm{Yb^{+}}$ in the state $5d^{2}D_{3/2}$ has been calculated
using relativistic coupled-cluster (RCC) theory. Indeed a comparison
of this property of a heavy ion like $\mathrm{Yb^{+}}$ with accurate
experimental data is a far more stringent test of RCC than the corresponding
comparison for Sr$^{+}$. The calculation of EQM for  $\mathrm{Yb^{+}}$ is 
computationally more demanding due to the presence of a large number of occupied
orbitals. In our calculation, the entire core has been excited. This leads
to a rapid proliferation in the number of cluster amplitude equations with the size of the
virtual space considered, and therefore a very large increase in the number of computations
necessary to determine these amplitudes. Obtaining convergence for the large number of 
cluster amplitude equations with an appropriate iterative 
method is a daunting task for heavy atomic systems like $\mathrm{Yb^{+}}$.
The calculations on such systems hence involve the combination of the power 
of the relativistic many-body theories with the state-of-the-art
high performance computational techniques.

An outline of the application of the RCC method to calculate atomic
electric quadrupole moments has already been presented in \cite{csur-PRL}.
The details of this theory have been discussed in several papers \cite{bishop,bartlett}.
Here we shall give the salient features of the method for completeness.
This paper is organized as follows : Section \ref{Theory} and Section \ref{compute} deal with
the theoretical methods we have employed and the details of our calculation respectively. 
In Section \ref{Results} we present our
results and compare with the available data. We have also discussed
the effects of different many-body contributions. Finally we conclude
in Section \ref{concl} and highlight the important findings of our
work.

\section{\label{Theory} Theoretical Methods}

\subsection{\label{RCC}Relativistic coupled-cluster theory for closed shell atoms}

We start with the $N$-electron closed-shell Dirac-Fock (DF) reference
state $\left|\Phi\right\rangle $, which is the Fermi vaccuum for the present formulation. 
In coupled-cluster (CC) theory
the exact wavefunction for the core sector in terms of this reference
state is given by, \begin{equation}
\left|\Psi\right\rangle =\exp(T)\left|\Phi\right\rangle ,\label{T-eqn}\end{equation}
 where $T$ is the cluster operator which takes into account the excitations
from the closed-shell core to the virtual orbitals. In singles and
doubles (SD) approximation, the cluster amplitude $T$ is written
as \begin{equation}
T=T_{1}+T_{2}=\sum_{ap}\left\{ a_{p}^{\dagger}a_{a}\right\} t_{a}^{p}+\frac{1}{4}\sum_{abpq}\left\{ a_{p}^{\dagger}a_{q}^{\dagger}a_{b}a_{a}\right\} t_{ab}^{pq};\label{T-single-part}\end{equation}
 $T_{1}$ and $T_{2}$ being the cluster amplitudes for single and
double excitations respectively and the curly brackets denote the {\emph normal ordering}
with respect to the Fermi vaccuum. This is known as coupled-cluster
with singles and doubles, namely CCSD. Here $t_{a}^{p}$ and $t_{ab}^{pq}$
are the corresponding single particle amplitudes and $a,b..(p,q..)$
stand for occupied (virtual) orbitals and $\left\{ \cdots\right\} $
denotes normal ordering with respect to the common reference state
(vacuum) $\left|\Phi\right\rangle $. For a one-valence one-dimensional model space
the label `v' is used to represent a valence orbital. In our approach we deal with
the normal ordered Hamiltonian which is defined as

\begin{equation}
\mathcal{H}\equiv H-\langle\Phi|H|\Phi\rangle=H-E_{DF},\label{Norm-Ham}\end{equation}
 where $E_{DF}$ is the Dirac-Fock energy.

\subsection{\label{OSCC-EA}Open shell coupled-cluster theory for single valence
system : Electron Attachment (OSCC-EA)}

To determine the wavefunctions for the open shell orbitals we employ
open-shell coupled-cluster method for electron attachment (OSCC-EA)
for the valence particle ($0h,1p$) sector. Using the scheme of electron
attachment (EA) we obtain the $(N+1)$-electron open shell system
as

\begin{equation}
\mathrm{Atom(0,0)}+e\longrightarrow\mathrm{Ion(0,1)}\,.\label{Fock-space scheme-EA}\end{equation}
 For a single valence system, we start with the reference state \begin{equation}
\left|\Phi_{v}^{N+1}\right\rangle \equiv a_{v}^{\dagger}\left|\Phi\right\rangle \label{Ref-Open}\end{equation}
 where $v$ denotes the valence orbital as mentioned in the previous section and the operator $a_{v}^{\dagger}$
represents creation of a particle in the valence space. The many-body
exact open-shell wavefunction for the $(N+1)$-electron open shell
system now becomes,

\begin{equation}
\left|\Psi_{v}^{N+1}\right\rangle =\exp(T)\left\{ \exp(S_{v})\right\} \left|\Phi_{v}^{N+1}\right\rangle ,\label{open-exp}\end{equation}
where the curly brackets denote the {\emph normal ordering} with respect to $|\Phi\rangle $.
 For a single valence system, the operator $\exp(S_{v})$ turns out
to be $(1+S_{v})$

\begin{equation}
\left|\Psi_{v}^{N+1}\right\rangle =\exp(T)\left\{ \left(1+S_{v}\right)\right\} \left|\Phi_{v}^{N+1}\right\rangle ,\label{open-exp-SV}\end{equation}
 with \begin{equation}
S_{v}=S_{1v}+S_{2v}=\sum_{v\neq p}\left\{ a_{p}^{\dagger}a_{v}\right\} s_{v}^{p}+\frac{1}{2}\sum_{bpq}\left\{ a_{p}^{\dagger}a_{q}^{\dagger}a_{b}a_{v}\right\} s_{vb}^{pq}\,.\label{S-EA}\end{equation}
 Here $S_{v}$ corresponds to the excitation operator in the valence
($v$) sector and and $s_{v}^{p}$ and $s_{vb}^{pq}$ are the singles
and doubles amplitudes respectively. The evaluation of the cluster
amplitudes are discussed elsewhere \cite{napp-IP}. Apart from singles
and doubles, only approximate triple excitations (CCSD(T)) have been
included. In this calculation, we have used OSCC-EA to obtain the
$5d\,^{2}D_{3/2}$ state of $^{171}{\rm Yb}^{+}$which is followed by property
calculations as given in sub-section \ref{Prop-calc}.

\subsection{\label{Prop-calc}Calculation of Expectation values}

The expectation value of any operator $O$ with respect to the exact
state is given by

\begin{eqnarray}
 & \left\langle O\right\rangle  & =\frac{\left\langle \Psi^{N+1}\right|O\left|\Psi^{N+1}\right\rangle }{\left\langle \Psi^{N+1}\right|\left.\Psi^{N+1}\right\rangle }\nonumber \\
 &  & =\frac{\left\langle \Phi^{N+1}\right|\left\{ 1+S^{\dagger}\right\} \bar{O}\left\{ 1+S\right\} \left|\Phi^{N+1}\right\rangle }{\left\langle \Phi^{N+1}\right|\left\{ 1+S^{\dagger}\right\} \exp(T^{\dagger})\exp(T)\left\{ 1+S\right\} \left|\Phi^{N+1}\right\rangle }.\label{prop-eqn}\end{eqnarray}
 where $\bar{O}=\exp(T_{c}^{\dagger})O\exp(T_{c})$ is the dressed
operator. The first few terms of the operators in the above expression
(Eq. (\ref{prop-eqn})) can be identified as $\bar{O}$, $\bar{O}S_{1}$,
$\bar{O}S_{2}$, $S_{1}^{\dagger}\bar{O}S_{1}$ etc. The corresponding
matrix elements are referred to as dressed Dirac-Fock (DDF), dressed
pair correlation (DPC) and dressed core polarization (DCP) respectively.
We use the term `dressed' because the operator $\bar{O}$ includes
the effects of certain core excitations, i.e., core-correlation effects.
In addition to the above, we can identify a few other terms which play a non-negligible
role in determining the correlation effects. One of those terms is
$S_{1}^{\dagger}\bar{O}S_{1}+c.c$ which we call the dressed higher
order pair correlation (DHOPC) since it directly involves the correlation
between a pair of electrons. Diagrams representing these terms
have already been presented in ref. \cite{csur-PRL} but we have nevertheless given them
here for clarity.

\begin{figure}[h]
\begin{centering}
\includegraphics{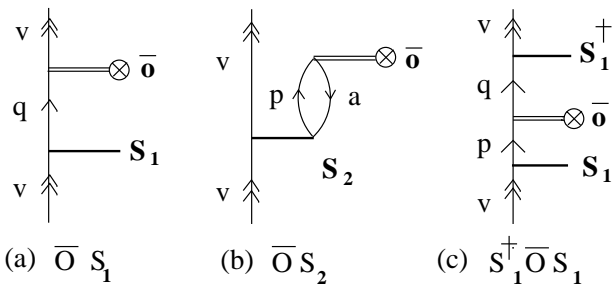}
\par\end{centering}

\caption{\label{obard}The diagrams (a) and (c) are subsets of dressed pair
correlation (DPC) diagrams. Diagram (b) is one of the direct dressed
core-polarization (DCP) diagram.}
\end{figure}

\subsection{\label{EQM-theory}Electric Quadrupole Moment}

The interaction of the atomic quadrupole moment with the external
electric-field gradient is analogous to the interaction of a nuclear
quadrupole moment with the electric fields generated by the atomic
electrons at the nucleus. In the presence of the electric field,
this gives rise to an energy shift by coupling with the gradient of
the electric field. Thus the treatment of atomic electric quadrupole moment
is analogous to its nuclear counterpart.

The quadrupole moment ${\bf \Theta}$ of an atomic state $|\Psi(\gamma,J,M)\rangle$
is defined as the diagonal matrix element of the quadrupole operator
with the maximum value $M_{J}$ and is expressed as

\begin{equation}
{\bf \Theta}=\left\langle \Psi(\gamma JFM_{F})\right|\Theta_{zz}\left|\Psi(\gamma JFM_{F})\right\rangle \,.\label{theta}\end{equation}
 Here $\gamma$ specifies the electronic configuration of the atoms
which distinguishes the initial and final states; $J$ is the total
angular momentum of the atom and $F$ is the summation of nuclear
and atomic angular momentum with $M_{F}$ its projection. The electric
quadrupole operator in terms of the electronic coordinates is given
by, \[
\Theta_{zz}=-\frac{e}{2}\sum_{j}\left(3z_{j}^{2}-r_{j}^{2}\right),\]
 where the sum is over all the electrons and $z$ is the coordinate
of the $j$th electron. To calculate the quantity we express the quadrupole
operator in its single particle form as \begin{equation}
\Theta_{m}^{(2)}=\sum_{m}q_{m}^{(2)}\label{theta-single}\end{equation}
 and the single particle reduced matrix element is expressed as \cite{grant-rad}

\begin{equation}
\left\langle j_{f}\right\Vert q_{m}^{(2)}\left\Vert j_{i}\right\rangle =\left\langle j_{f}\right\Vert C_{m}^{(2)}\left\Vert j_{i}\right\rangle \int dr~r^{2}\left(\mathcal{P}_{f}\mathcal{P}_{i}+\mathcal{Q}_{f}\mathcal{Q}_{i}\right).\label{quad-eq}\end{equation}
 In Eq.($\ref{quad-eq}$), the subscripts $f$ and $i$ correspond
to the final and initial states respectively; $\mathcal{P}$ and $\mathcal{Q}$
are the radial part of the large and small components of the single
particle Dirac-Fock wavefunctions respectively and $j_{i}$ is the
total angular momentum for the $i$th electron. The angular factor
is given by \begin{eqnarray}
\left\langle j_{f}\right\Vert C_{m}^{(k)}\left\Vert j_{i}\right\rangle = &  & (-1)^{(j_{f}+1/2)}\sqrt{(2j_{f}+1)}\sqrt{(2j_{i}+1)}\nonumber \\
 &  & \times\left(\begin{array}{ccc}
j_{f} & 2 & j_{i}\\
-1/2 & 0 & 1/2\end{array}\right)\pi(l,k,l^{\prime})\label{ang}\end{eqnarray}
 where \[
\pi(l,k,l^{\prime})=\left\{ \begin{array}{c}
\begin{array}{cc}
1 & \mathrm{if}\: l+k+l^{\prime}\,\,\mathrm{even}\\
0 & \mathrm{otherwise}\end{array}\end{array}\right.\]
 $l$ and $k$ being the orbital angular momentum and the rank respectively.

Finally using the Wigner Eckart theorem we define the electric quadrupole
moment in terms of the reduced matrix elements as \begin{equation}
\left\langle j_{f}\right|\Theta_{m}^{(2)}\left|j_{i}\right\rangle =(-1)^{j_{f}-m_{f}}\left(\begin{array}{ccc}
j_{f} & 2 & j_{i}\\
-m_{f} & 0 & m_{f}\end{array}\right)\left\langle j_{f}\right\Vert \Theta^{(2)}\left\Vert j_{i}\right\rangle \label{wig-eck}\end{equation}

\section{\label{compute} Computational Details}

This calculation is performed in the following steps : The first step
being the generation of single particle basis for $\mathrm{Yb^{++}}$
using the Gaussian basis set expansion. This is followed by the generation
of the coupled cluster amplitudes ($T$) for the closed-shell $\mathrm{Yb^{++}}$
system. In the next step, the virtual orbitals $6s$ and $5d_{3/2}$
are generated using the open-shell coupled cluster method for electron
attachment (OSCC-EA). This is followed by the property calculations
as given in subsection \ref{Prop-calc}.

The orbitals used in the present work are generated by kinetically
balanced finite basis set expansion (FSBE) of Gaussian type orbitals
(GTO) \cite{napp-fbse} \begin{equation}
F_{i,k}(r)=r^{k}\exp(-\alpha_{i}r^{2}),\label{comp-1}\end{equation}
 with $k=0,1,2\cdots$ for $s,p,d,\cdots$ type functions, respectively.
The exponents are determined by the even tempering condition \cite{even-tem}
\begin{equation}
\alpha_{i}=\alpha_{0}\beta^{i-1}.\label{comp-2}\end{equation}
 The starting point of the computation is the generation of the Dirac-Fock
(DF) orbitals \cite{napp-fbse} which are defined on a radial grid
of the form \begin{equation}
r_{i}=r_{0}\left[\exp(i-1)h-1\right]\label{comp-3}\end{equation}
 with the freedom of choosing the parameters $r_{0}$ and $h$. All
DF orbitals are generated using a two parameter Fermi nuclear distribution
\begin{equation}
\rho=\frac{\rho_{0}}{1+\exp((r-c)/a)},\label{fermi-nucl}\end{equation}
 where the parameter $c$ is the half charge radius and $a$ is related
to skin thickness, defined as the interval of the nuclear thickness
in which the nuclear charge density falls from near one to near zero.
Table \ref{gauss-table} contains the information about the basis
functions used in the calculation to determine the electric quadrupole
moment of $5d\,^{2}D_{3/2}$ state of $^{171}{\rm Yb}^{+}$.

\begin{table}[h!]

\caption{\label{gauss-table}No. of basis (NB) functions used to generate
the even tempered Dirac-Fock orbitals and the corresponding value
of $\alpha_{0}=\alpha\times10^{-5}$ and $\beta$ used. NP and NH
stand for number or particles and number of holes respectively. }

\begin{centering}
\begin{tabular}{llllllllll}
\hline 
&
$s_{1/2}$&
$p_{1/2}$&
$p_{3/2}$&
$d_{3/2}$&
$d_{5/2}$&
$f_{5/2}$&
$f_{7/2}$&
$g_{7/2}$&
$g_{9/2}$\tabularnewline
\hline
\hline 
NB&
38&
35&
35&
25&
25&
25&
25&
20&
20 \tabularnewline
$\alpha$ &
305&
325&
325&
335&
335&
315&
315&
345&
345 \tabularnewline
$\beta$&
2.106 &
2.116 &
2.116 &
2.316 &
2.316 &
2.216 &
2.216 &
2.135 &
2.135 \tabularnewline
NP&
7 &
8 &
8 &
7 &
7 &
7&
7&
8&
8 \tabularnewline
NH&
5&
4&
4&
2&
2&
1&
1&
0&
0 \tabularnewline
\hline
\end{tabular}
\par\end{centering}
\end{table}

\section{\label{Results} Results and Discussion }

The contribution of the important physical effects to the electric quadrupole moment of
$^{171}{\rm Yb}^{+}$ is given in Table.\ref{contrbn}. The total value
that we have obtained for this quantity $\Theta_{3/2}(\mathrm{calculated})=2.157ea_{0}^{2}$
is within the error bounds of the measured value $\Theta_{3/2}
(\mathrm{measured})=(2.08 \pm 0.11) ea_{0}^{2}$
\cite{Yb+_Q-exp}. It would be instructive to compare our present
calculation based on the RCC theory with a previous calculation performed
by Itano using the relativistic configuration interaction (RCI) method
\cite{Itano-pra}. The value obtained by Itano is $\Theta_{3/2}
=2.174 ea_{0}^{2}$ \cite{Itano-pra}.
The RCI calculation of Itano uses a multiconfiguration
Dirac-Fock (MCDF) extended optimized level (EOL) orbital basis.
The configurations included in the latter calculation
constitute a subset of the configurations in our calculation. In particular,
they correspond to the correlation effects arising from the single
and double excitations from the core and the valence, i.e., terms
involving $T_{1}$, $T_{2}$, $S_{1}$ and $S_{2}$ in our RCC calculation.
In our calculation, the above effects have been included to all orders.
The virtual orbitals considered by Itano were $\{(6-10)s,(6-10)p,(6-10)d,(5-7)f,(5-6)g,6h\}$.
The calculation carried out by Itano involves 2 main steps. In the first step, 
the SPOs are obtained by minimizing an energy functional in a limited orbital space 
in the framework of MCDF-EOL. The second step involves a fairly large RCI calculation to account for
the correlation effects. Itano's calculation incorporates the valence-core
correlation (single and the double excitations) arising from the \{5d\}
and \{4f,5s,5p\} shells and single excitations from \{4s,4p,4d,3d\}
(core-core correlations). Inspite of these differences, the results
are in good agreement.

For each symmetry we have considered more virtual orbitals than Itano.
In addition, we have considered single and double excitations from
all the core orbitals where Itano has considered only single and double
excitations from $5d$ and $\{4f,5s,5p\}$ core orbitals, but not more than
one single core excitation at a time.

From Table. \ref{contrbn}, we see that the dressed Dirac-Fock
contribution is the largest and it is a substantial fraction of the
total EQM, inspite of the large number of core-valence excitations. 
The second largest contribution
is from the DPC effects and third being DCP effects. DPC effects arise
from the terms like $\overline{O}S_{1}$ and $S_{1}^{\dagger}\overline{O}S_{1}$.
Sub-section \ref{Prop-calc} gives the details of these different
many-body terms. $S_{1}$ is an operator of rank 0 and the valence
orbital in $^{171}{\rm Yb}^{+}$ is a $5d_{3/2}$ orbital and hence it
excites the valence electron to a virtual orbital of the same parity and 
angular momentum
giving a large contribution through the virtual $d$ orbitals. 
Clearly the effects of the valence $5d_{3/2}$ excitations are the
most important in the case of $\mathrm{^{171}Yb^{+}}$. It is interesting to note that
though $\mathrm{^{171}Yb^{+}}$ has more filled shells than for $\mathrm{^{88}Sr^{+}}$,
the contribution of the DDF, DPC, DCP terms follow the same trend
for $\mathrm{^{171}Yb^{+}}$ and $\mathrm{^{88}Sr^{+}}$. The core-virtual
electric quadrupole excitations/deexcitations contributing to the DCP diagrams involve excitations
from $f$ - $f$ or $d$ - $d$ or $f$ - $p$, etc orbitals. The DCP diagrams
involve the contributions from the $\bar O$ and the $S_2$ matrix elements.
Though the $\bar O$ matrix element could be large in some cases, the contribution of the matrix element of the
product $\bar O S_2$ turns out to be two orders of magnitude smaller than
that of the DDF term, inspite of the possibility of large number of 
core-virtual excitations. Similar trends were observed for
the EQM of $\mathrm{^{88}Sr^{+}}$. Table \ref{gauss-table} has more details about
the virtual orbitals and the active space. From Table. \ref{contrbn}, we find that
the contributions of the terms DDF, DPC, DCP and DHOPC to the final
value are 119$\%$, -14$\%$, -2.2$\%$, 1.2$\%$ respectively for
$\mathrm{^{88}Sr^{+}}$ and are 116$\%$, -13.3$\%$, -1.3$\%$, 1.34
$\%$ respectively for $\mathrm{^{171}Yb^{+}}$. Also, the total contribution
of DPC, DCP and DHOPC to DDF is -1.1 $\%$. DPC is $\sim11\%$ of
the DDF. DCP is $\sim1\%$ of DDF for the case of $\mathrm{^{171}Yb^{+}}$. 
It has been observed that this trend is similar for $\mathrm{^{88}Sr^{+}}$
even though $\mathrm{^{171}Yb^{+}}$ has a larger core.
\begin{table}

\caption{\label{contrbn}Contributions from the electric quadrupole moment
(in $ea_{0}^{2}$) of the $5d\,^{2}D_{3/2}$ state of $\mathrm{^{171}Yb^{+}}$
and the $4d\,^2D_{5/2}$ state of $\mathrm{^{88}Sr^{+}}$,
corresponding to different many-body effects in the CCSD calculation.
The terms like DDF, DCP, DPC, DHOPC are explained in the text. }

\begin{centering}
\begin{tabular}{lllllll}
&
&
&
&
&
&
\tabularnewline
\hline 
Ion &
DDF&
DPC&
DCP&
DHOPC&
Total &
Expt. \cite{Yb+_Q-exp}\tabularnewline
\hline
\hline 
&
&
&
&
&
&
\tabularnewline
${\rm Yb}^{+}$ &
2.500 &
-0.287 &
-0.0280 &
0.029 &
2.157 &
2.08$\pm$0.11\tabularnewline
$Sr^{+}$ &
3.496 &
-0.4306 &
-0.0642 &
0.0353 &
2.94 &
2.6$\pm $ 0.3
\tabularnewline
\hline\hline&
&
&
&
&
&
\tabularnewline
\end{tabular}
\par\end{centering}
\end{table}

\section{\label{concl} Conclusions}

In summary, we have used relativistic coupled-cluster theory to calculate
the electric quadrupole moments (EQM) of the $5d\,^{2}D_{3/2}$ 
state of $\mathrm{^{171}Yb^{+}}$.
Our determination of EQM of the $5d\,^{2}D_{3/2}$state of $\mathrm{^{171}Yb^{+}}$
is within the experimental limits. It highlights the ability of RCC to capture
the interplay between the relativistic and the correlation effects
in heavy single valence ions. We have also determined the various
leading many-body effects arising in this calculation. To our knowledge
this calculation yields the most accurate theoretical value of EQM
of the $5d\,^{2}D_{3/2}$state of $\mathrm{^{171}Yb^{+}}$to date.
It is a useful theoretical supplement to
the experimental search for optical frequency standards. 

\begin{acknowledgments}
This work is partially supported by BRNS project no. 2002/37/12/BRNS.
RKC acknowledges the Department of Science and Technology, India (grant
SR/S1/PC-32/2005). 
\end{acknowledgments}

\end{document}